\documentclass[twocolumn,showpacs,preprintnumbers,amsmath,amssymb]{revtex4}

\usepackage{graphicx}
\usepackage{dcolumn}
\usepackage{latexsym}
\usepackage{bm}

\begin{document}

\title{A lattice calculation of the pion form factor with Ginsparg-Wilson-type 
fermions}

\author{Stefano Capitani}
\email{stefano.capitani@uni-graz.at}
\author{Christof Gattringer}
\email{christof.gattringer@uni-graz.at}
\author{C. B. Lang}
\email{christian.lang@uni-graz.at}
\affiliation{{\rm(for the Bern-Graz-Regensburg (BGR)
collaboration)}\vspace{1mm}\\ Institut f\"ur Physik, FB Theoretische Physik\\
Universit\"at Graz, A-8010 Graz, Austria}

\date{November 18, 2005}

\begin{abstract}
Results for Monte Carlo calculations of the electromagnetic vector and
scalar form factors of the pion in a quenched simulation are presented.
We work with two different lattice volumes up to a spatial size of 2.4~fm
at a lattice spacing of 0.148~fm. The pion form factors in the space-like
region are determined for pion masses down to 340 MeV. 
\end{abstract}

\maketitle

\section{Introduction}

The ab-initio determination of hadronic properties is an important task of
lattice QCD calculations. Following the impressive progress in the
determination of the hadron mass spectrum, one went on to more advanced
calculations, such as the determination of matrix elements. An important
example is the electromagnetic pion form factor 
\cite{MaSa88,DrWoWi89,Ne04,HeKoLa04,BoEdFl05,AbLe05,Br05}.

With the successful implementation of chiral symmetry, based on the
Ginsparg-Wilson (GW) relation \cite{GiWi82}, also results at smaller
quark masses are feasible. In addition to an exact implementation
\cite{NaNe93a95,Ne9898a}, further Dirac operators have been suggested
that fulfill the GW condition in an approximate way, among them the
domain-wall fermions \cite{Ka92,FuSh95}, the perfect fermions \cite{HaNi94} 
and the chirally improved (CI) fermions \cite{Ga01,GaHiLa00}.  

Dynamical calculations at small quark masses are certainly the
ultimate goal. Currently such calculations are still
prohibitively expensive but one can try to improve in two ways:
either one works with dynamical fermions in a conventional
formulation or one tries to approach smaller quark masses with
GW type operators in the quenched approximation. In our
present contribution we focus on the latter.

We present here a first-principles calculation for off-forward lattice 
matrix elements of operators that measure the vector and scalar form 
factors of the pion. We work with the CI fermions, which constitute 
an approximate solution of the Ginsparg-Wilson relation and thus have 
improved chiral properties \cite{GaGoHa03a}. In our calculation we consider 
comparatively low pion masses down to 340 MeV, which is likely 
in the domain where chiral perturbation theory (ChPT) can be applied.
Another convenient feature of chirally improved fermions is that the 
renormalization of the local current $Z_V$ is quite close to 1
\cite{GaGoHu04}.

Preliminary results of our calculations have been presented in 
Ref.~\cite{CaGaLa05a}. The quenched results for the vector form factor at
the available transferred momenta lie close to the experimental curves.
From our data for the form factors we can derive the charge radius of the
pion as well as its scalar radius. The former comes out smaller than
experiment, due to the too large value of the $\rho$-meson mass. The
scalar radius appears to be sensitive to quenching and possibly to the
contributions of the disconnected diagrams, which we do not include in our
calculations. 

The pion form factor in the quenched approximation has been studied for 
Wilson type fermions \cite{MaSa88,DrWoWi89,HeKoLa04,BoEdFl05,Br05}, for
GW type fermions \cite{Ne04} and for twisted mass fermions \cite{AbLe05}. 
There are few results for dynamical fermions: for Wilson-type fermions in 
Refs.\ \cite{HaAoFu05,Br05}, and a study with domain-wall valence quarks 
and dynamical (MILC) Asqtad fermions \cite{BoEdFl05} 
(and further references therein).

\section{Pion form factors}

The vector form factor $F_\pi$ is defined by
\begin{equation}
\label{eq1}
\langle \pi^+ (\mathbf{p}_f) |\, V_\mu \,| \pi^+ (\mathbf{p}_i) 
\rangle_\textrm{\scriptsize cont}
= (p_f+p_i)_\mu \, F_\pi (Q^2)\;,
\end{equation}
where $Q^2 = (p_f-p_i)^2\equiv -t$ is the space-like invariant momentum 
transfer squared, and 
\begin{equation}
V_\mu = \frac{2}{3} \; \overline{u} \,\gamma_\mu \,u 
      - \frac{1}{3} \; \overline{d} \,\gamma_\mu \,d
\end{equation}
is the vector current.

The pion vector form factor is an analytic function of $t$. Its space-like 
values are determined from the values on the boundary of its analyticity 
domain, i.e., the cut along the positive real $t$-axis. Although it starts at 
$t=4 \,m_e^2$, significant contributions only come from the hadronic region, 
starting with the two-pion threshold at $4\,m_\pi^2$. Along the first part of 
the cut, until inelastic channels become important, due to unitarity the phase
shift is essentially the p wave phase shift of the elastic two-pion channel. 
That region is dominated by the $\rho$-meson resonance. 
This feature has been exploited by various dispersion relation
representations \cite{GoSa68,HeLa81,TrYn02,Le02}.

Although there are further inelastic contributions from the four-pion channel,
these remain tiny until the $\pi\omega$ channel opens. There is also a small
contribution from the isoscalar $\omega$, coupling through higher-order
electromagnetic interactions. All these contributions show small effects on 
the near space-like region and can be safely neglected within the 
accuracy attained in our work.

The dominant contribution of the vector meson resonances has inspired the 
vector meson dominance (VMD) model, where the form factor in the space-like 
domain is approximated by a sum of poles in the time-like region,
\begin{equation}
F_\pi^{VMD}=\sum_V \frac{f_{V\pi\pi}}{f_{V\gamma}}\frac{m_V^2}{m_V^2-t}\;.
\end{equation}
In first approximation the coefficients are related to the coupling
$f_{V\pi\pi}$ of the vector meson to the two-pion state and its coupling to the
photon, $f_{V\gamma}$. 

In QCD the large $Q^2$ behavior in leading order \cite{FaJa79} is given by
\begin{equation}
F_\pi (Q^2) \sim \frac{8\,\pi \,\alpha_s (Q^2) \,f_\pi^2}{Q^2}\;,
\end{equation}
with $\alpha_s$ the running coupling constant (logarithmic in the argument) and
$f_\pi$ the pion decay constant. In the $Q^2$-region accessible to us the
logarithm is not identifiable and the VMD behavior provides a good 
approximation.

Due to electric charge conservation one has $F_\pi(0)=1$. The mean charge 
radius squared is defined through
\begin{eqnarray}
\label{defvectorradius}
F_\pi (Q^2) & =& 1 - \frac{1}{6} \, \langle r^2 \rangle_\textrm{\scriptsize v} 
\, Q^2 + {\cal O}\big(Q^4\big)\nonumber\\
&\rightarrow &
\langle r^2 \rangle_\textrm{\scriptsize v} \equiv 6\,d F_\pi(t)/dt\vert_{t=0} 
\;.
\end{eqnarray}
The current PDG average for its value is $0.45(1)\;\textrm{fm}^2$
\cite{PDG04}.

In the simplest VMD model with just the leading $\rho$-resonance one has 
(due to normalization) $f_{\rho\pi\pi}=f_{\rho\gamma}$ and
\begin{eqnarray}
\label{eq:rvmd}
F_\pi (Q^2) &=& {m_\rho^2}/({m_\rho^2-t})\nonumber\\
&\rightarrow &
\langle r^2 \rangle_\textrm{\scriptsize v,VMD}=\frac{6}{m_\rho^2}\approx 0.39 
\;\textrm{fm}^2\;.
\end{eqnarray}

The scalar form factor $\Gamma_\pi$ is given by the matrix element of the 
scalar operator, i.e., 
\begin{equation}\label{scalarmatrixelement}
\langle \pi^+ (\mathbf{p}_f) | \,m_u\,\overline{u}\, u + m_d\,\overline{d} 
\,d\,| \pi^+ (\mathbf{p}_i) \rangle = \Gamma_\pi (Q^2)\;.
\end{equation}
Within chiral perturbation theory the scalar form factor
at $Q^2=0$ is the so-called sigma term which behaves like
$\Gamma_\pi (0) \sim M_\pi^2$ near zero momentum transfer. 
The scalar radius squared $\langle r^2 \rangle_s$ can be obtained from
\begin{equation}\label{defscalarradius}
\frac{\Gamma_\pi (Q^2)}{\Gamma_\pi (0)} =
1 - \frac{1}{6} \, \langle r^2 \rangle_s \, Q^2 + {\cal O}\big(Q^4\big)\;.
\end{equation}
For a detailed discussion of chiral perturbation theory in this context see 
\cite{AnCaCo04}.

\begin{figure}[t]
\begin{center}
\includegraphics*[width=50mm]{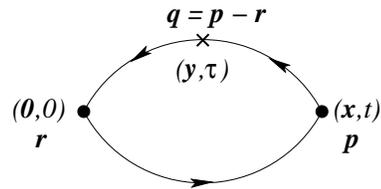}
\end{center}
\caption{Schematic diagram for the matrix element (\ref{eq:matele}).}
\label{fig_matele}
\end{figure}

\section{Strategy}\label{sec:strategy}

In order to compute the form factors, we need to evaluate off-forward matrix 
elements (at several transferred momenta), i.e., the expectation values of
\begin{equation}
{\mathrm{Tr}}\left[
\sum_{\mathbf{y};\,y_0=\tau}\!\!\mathrm{e}^{i\, \mathbf{q}\cdot\mathbf{y}} 
S(0,y) \; O \!\!
\sum_{\mathbf{x};\,x_0=t}\!\!\mathrm{e}^{-i\, \mathbf{p}\cdot\mathbf{x}} 
S(y,x) \gamma_5 S(x,0) \gamma_5 \right] \;,
\label{eq:matele}
\end{equation}
where $S(y,x)$ is the quark propagator from $x$ to $y$ and $O$ denotes the
operator inserted at $y$. We use the notation $\mathbf{p}\equiv\mathbf{p}_f$, 
$\mathbf{r}\equiv\mathbf{p}_i$ and denote the momentum transfer as
$\mathbf{q}=\mathbf{p}-\mathbf{r}$ (cf. Fig.\ \ref{fig_matele}).

The necessary matrix elements have been evaluated by using the sequential 
source method. The matrix element is then written as 
\begin{equation} 
\mathrm{Tr} \, \left[ 
\sum_{\mathbf{y};y_0=\tau} \, e^{i\, \mathbf{q}\cdot\mathbf{y}} \, S(0,y) 
\; O \; \Sigma(y,0) \, \gamma_5 \right] \;.
\end{equation}
The sequential propagator is defined as 
\begin{equation}
\Sigma(y,0) = \sum_{\mathbf{x};x_0=t} \, 
e^{-i\, \mathbf{p}\cdot\mathbf{x}} \, S(y,x) \, \gamma_5 \, S(x,0)\;,
\end{equation} 
and can be easily computed by an additional inversion of the Dirac operator
$D$ for each choice of the final momentum $\mathbf{p}$,
\begin{equation}
\sum_y D(z,y) \, \Sigma(y,0) =
e^{-i\, \mathbf{p}\cdot\mathbf{z}} \, \gamma_5 \, S(z,0) 
\Big|_{z_0=t}\;.
\end{equation}
Changing the properties of the sink requires the computation of new sequential
propagators, and so simulating several final momenta, different field 
interpolators, or a different smearing for the sink rapidly becomes rather
expensive. For this reason we have limited ourselves to only one value of
the final momentum. An alternative version of the sequential source
method, which builds the sequential propagator from the two propagators
that sandwich the operator, would allow more values of the final momentum
for free, but on the other hand it would require a new inversion of the
Dirac operator for every new type of operator and for each different
value of the momentum transfer. This is much less convenient in our
situation because we consider both the vector and scalar currents and want
to obtain the corresponding form factors for several values of the
momentum transfer.

We extract the physical matrix elements by computing ratios of 3-point and
2-point correlators:
\begin{widetext}
\begin{equation}
R(t,\tau;\mathbf{p},\mathbf{q}) = 
\frac{\langle P(t;\mathbf{p})\,O(\tau;\mathbf{q})\,\overline{P}(0;\mathbf{r}) 
\rangle
}{\langle P(t;\mathbf{p})\,\overline{P}(0;\mathbf{p}) \rangle} \,
\sqrt{ \, 
    \frac{\langle P(t;\mathbf{p})\,\overline{P}(0;\mathbf{p})\rangle \,
          \langle P(\tau;\mathbf{p})\,\overline{P}(0;\mathbf{p})\rangle \,
          \langle P(t-\tau;\mathbf{r})\,\overline{P}(0;\mathbf{r})\rangle }{
          \langle P(t;\mathbf{r})\,\overline{P}(0;\mathbf{r})\rangle \,
          \langle P(\tau;\mathbf{r})\,\overline{P}(0;\mathbf{r})\rangle \,
          \langle P(t-\tau;\mathbf{p})\,\overline{P}(0;\mathbf{p})\rangle }}\;,
\label{eqcorr}
\end{equation}
\end{widetext}
where $P=\overline{u} \,\gamma_5\,d$ is our pseudoscalar interpolator. We
keep the sink fixed at time $t$ and vary the timeslice $\tau$ where the
operator $O$ sits (scanning a range of timeslices).

The ratio (\ref{eqcorr}) eliminates the exponential factors in the time 
variable which are present if one considers $\langle P(t;\mathbf{p})\,
O(\tau;\mathbf{q})\,\overline{P}(0;\mathbf{r}) \rangle$ alone. As
a consequence, $R(t,\tau;\mathbf{p},\mathbf{q})$ exhibits two plateaus in
$\tau$: $0 \ll \tau \ll t$ and $t \ll \tau \ll T$. In the case in which
the sink is put at $t=T/2$ the ratio is antisymmetric in $\tau-T/2$. 

In the Monte Carlo simulations the momentum projections of the sink onto 
non-zero $\mathbf{p}$ and the operator onto non-zero momentum transfer
$\mathbf{q}$ lead to significant statistical fluctuations due to the worse
signal-to noise ratio. In particular, for many configurations the 2-point
correlators become negative on timeslices near the symmetry point
$t=T/2$. A reasonable way out of this inconvenient situation is to choose
the sink to sit at a timeslice smaller than $T/2$, because in this region
the 2-point correlators have much larger values, and moreover their
relative errors are smaller. We have thus put the sink at timeslice $t=7$,
while the source remained at timeslice $t=0$. The matrix elements are
finally determined (as in the case $t=T/2$) by combining the two plateau
values on either side of the sink, 
$R_\textrm{\scriptsize lhs}\pm R\textrm{\scriptsize rhs}$, where the
relative sign depends on the parity properties of the operators. This
sign is negative for the vector and positive for the scalar form factor.

The term under the square root in Eq.\ (\ref{eqcorr}) becomes trivial for
$\mathbf{q}=(0,0,0)$. In that case the ratio becomes
\begin{equation}
\frac{\langle P(t;\mathbf{p})\,O(\tau;\mathbf{q}=
\mathbf{0})\,\overline{P}(0;\mathbf{r}=\mathbf{p}) \rangle
}{\langle P(t;\mathbf{p})\,\overline{P}(0;\mathbf{p}) \rangle}
\end{equation}
and is sufficient to eliminate the exponentials.

\begin{table}[t]
\caption{Values of the non-zero momentum transfers.}
\begin{ruledtabular}
\begin{tabular}{ccccc}
size & $\sqrt{2}\,p_0$ & $2\,p_0$ &$\sqrt{6}\,p_0$ & $2\sqrt{2}\,p_0$ \\
\hline
~\vspace{-8pt}\\
$16^3\times 32$ & 0.739~GeV & 1.045~GeV & 1.280~GeV & 1.478~GeV \\
$12^3\times 24$ & 0.986~GeV & 1.394~GeV & 1.707~GeV & 1.971~GeV \\
\end{tabular}
\end{ruledtabular}
\label{tab:momenta}
\end{table}

We choose momenta $|\mathbf{p}_f | = | \mathbf{p}_i |$, which implies 
$E_{f} = E_{i}$ so that the transferred 4-momentum is then given by 
$Q^2 = |\mathbf{q}|^2$. In this way \cite{HeKoLa04} one achieves for the
electromagnetic form factor a cancellation of the kinematical factors in
(\ref{eq1}) and
\begin{equation}
\langle \pi (\mathbf{p}_f) | V_\mu | \pi (\mathbf{p}_i) 
\rangle_\textrm{\scriptsize latt}
= \frac{1}{2\,\sqrt{E_f E_i}}
\langle \pi (\mathbf{p}_f) | V_\mu | \pi (\mathbf{p}_i) 
\rangle_\textrm{\scriptsize cont}\;.
\end{equation}
Indeed, with this choice of momenta, and using the $\mu =4$ component of 
the vector current, the overall factor becomes 
\begin{equation}
\frac{E_f + E_i}{2\sqrt{E_f E_i}} = 1 \;.
\label{eqsimplification}
\end{equation}
This cancellation is unfortunately no longer possible in the calculation 
of the scalar form factor, where a remaining multiplication by the
quantity $2\,\sqrt{E_f \,E_i}$ is still needed. 

In our calculations the pion source is located at the origin and the sink 
has a non-zero 3-momentum which is fixed to $\mathbf{p}_f=(1,1,0)\,p_0$,
where $p_0=2\,\pi/ (a\,L)$ denotes the smallest non-vanishing spatial
momentum. The momenta of the current are then chosen such that implicitly
\begin{equation}
\label{eq:momcondition} |\mathbf{p}_f| = |\mathbf{p}_i|= \sqrt{2}\, p_0 
\end{equation}
is verified. We use twelve different 3-momenta for the current, 
$(0,0,0)$, $(0,1,\pm 1)$, $(1,0,\pm 1)$, $(2,0,0)$, $(0,2,0)$, $(2,1,\pm
1)$, $(1,2,\pm 1)$, $(2,2,0)$; these are all the possible choices that
obey the condition (\ref{eq:momcondition}). This setup then gives rise to
four non-zero (and equidistant) values for the square of the momentum
transfer, $Q^2 = 2\,n \,p_0^2$ for $n=0,\,1,\,2,\,3,\,4$ (see
Table~\ref{tab:momenta}). Momentum conservation finally defines the 
twelve corresponding values for the 3-momentum of the source, which all
have the same module.

We could also consider larger values of the module of the initial
momentum, since this would generate further values of $Q^2$ while still
using the same sequential propagators through the sink. In this case,
however, the statistical errors become much larger, and moreover we could
not use Eq.\ (\ref{eqsimplification}) anymore. A change of the momentum at
the sink would require the expensive computation of a new set of
sequential propagators.

\section{Simulation parameters}

We use the chirally improved Dirac operator
\cite{Ga01,GaHiLa00,GaGoHa03a}, which constitutes an approximate
Ginsparg-Wilson operator. The gauge action is the L\"uscher-Weisz tadpole
improved action at $\beta=7.9$ \cite{LuWe85}, corresponding to a lattice
spacing of 0.148~fm (determined from the Sommer parameter in
\cite{GaHoSc02}). Here we do not study the scaling behavior. However,
although we cannot estimate the discretization effects, we expect that our
results have only small $O(a)$ contributions, because chirally improved
fermions are Ginsparg-Wilson-like fermions. This has been observed in the
study of the hadron spectrum \cite{GaGoHa03a} and other hadron properties
\cite{GaHuLa05}.

The pion interpolators are constructed from Jacobi-smeared quark sources
and sinks (with $\kappa=0.21$ and $N=18$), which improves the signal for
the ground state.

We performed simulations on two volumes: $12^3\times 24$ (200
configurations for each mass) and $16^3\times 32$ (100 configurations for
each mass). This corresponds to spatial lattice sizes of $a\,L \sim
1.8\;$fm and $a\,L \sim 2.4\;$fm respectively. 

We work at several values of the bare quark mass and the corresponding
meson masses are given in Table~\ref{tab:mesonmasses} (taken from
\cite{GaGoHa03a}). The jackknife method is used for estimating the
statistical errors of our results. 

\begin{table}[t]
\caption{Pion and rho masses at the various quark masses simulated
(taken from \cite{GaGoHa03a}).}
\label{tab:mesonmasses}
\begin{ruledtabular}
\begin{tabular}{c|ccc|cc}
& $16^3\times 32$ & $16^3\times 32$ & $16^3\times 32$ 
& $12^3\times 24$ & $12^3\times 24$ \\
\hline
~\vspace{-8pt}\\
$m$      &  0.02  $a^{-1}$   &  0.04  $a^{-1}$   &  0.06  $a^{-1}$   
         &  0.04  $a^{-1}$   &  0.06  $a^{-1}$   \\ 
$m_\pi$  &  342~MeV  &  471~MeV  &  571~MeV  &  474~MeV  &  575~MeV  \\ 
$m_\rho$ &  845~MeV  &  895~MeV  &  941~MeV  &  952~MeV  &  976~MeV  \\
\end{tabular}
\end{ruledtabular}
\end{table}

\section{Results}

\begin{figure}[t]
\begin{center}
\includegraphics*[width=87mm]{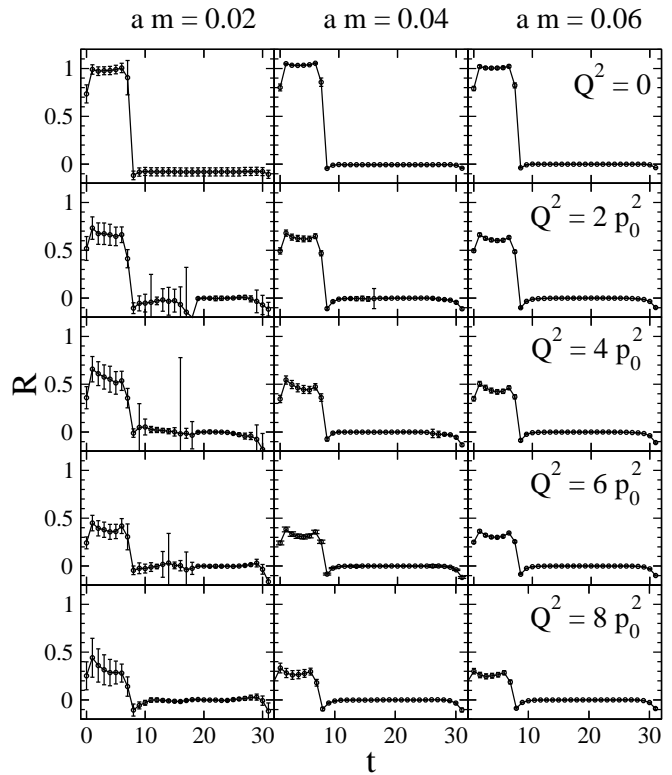}
\end{center}
\caption{The ratio (\ref{eqcorr}) leading to the vector form factor for all
quark masses and transfer momenta considered (lattice size $16^3\times 32$).}
\label{fig_rat_vect}
\end{figure}

In order to illustrate the quality of the results, we show in Figs.
\ref{fig_rat_vect} and \ref{fig_rat_scal} the time dependence of the
ratio (\ref{eqcorr}). Fitting the central points of the plateaus to a
constant then leads to the values for the form factors. 

\begin{table}[b]
\caption{Values of the pion vector form factor (lattice size $16^3\times32$,
$a=0.148$~fm) obtained for different quark mass values.}
\label{tab:fpi_vect}
\begin{ruledtabular}
\begin{tabular}{cccc}
~\vspace{-8pt}\\
  $ Q^2$[GeV]  &     $a\, m$=0.02&  $a \,m$=0.04&   $a \,m$=0.06\\
\hline
  0    &  1.01(1)  & 0.996(3)  &  0.963(13)\\
 0.546 &  0.64(15) & 0.596(42) &  0.580(23)\\
 1.093 &  0.54(14) & 0.434(42) &  0.409(24)\\
 1.639 &  0.36(12) & 0.296(34) &  0.289(20)\\
 2.186 &  0.29(14) & 0.252(43) &  0.239(25)\\
\end{tabular}
\end{ruledtabular}
\end{table}

\begin{figure}[t]
\begin{center}
\includegraphics*[width=87mm]{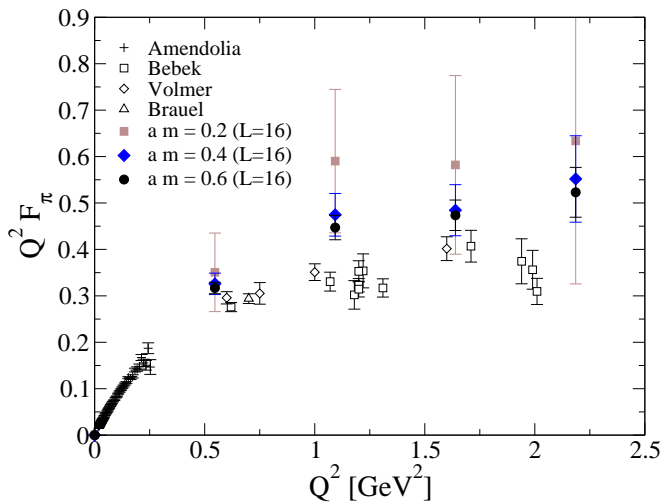} 
\end{center}
\caption{Our results for the pion vector form factor $Q^2\,F_\pi$ compared with
results from experiments (lattice size $16^3\times 32$).}
\label{fig_fpi_q2}
\end{figure}

\begin{figure}[t]
\begin{center}
\includegraphics*[width=87mm]{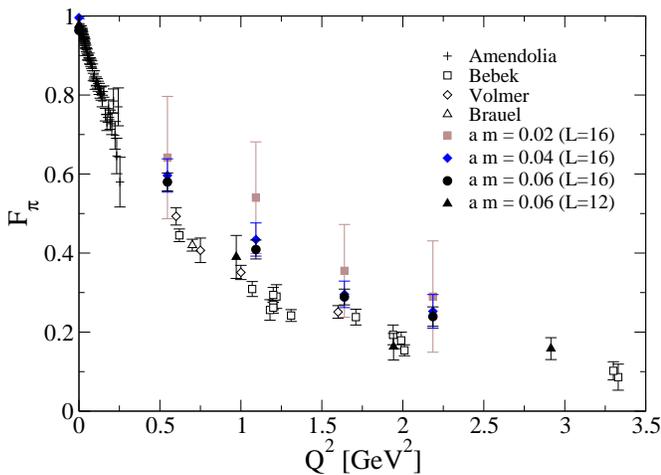} 
\end{center}
\caption{Our results for the pion vector form factor $F_\pi$ compared with
results from experiments (lattice sizes $16^3\times 32$ and $12^3\times 24$).}
\label{fig_fpi_all}
\end{figure}

\begin{figure}[t]
\begin{center}
\includegraphics*[width=87mm]{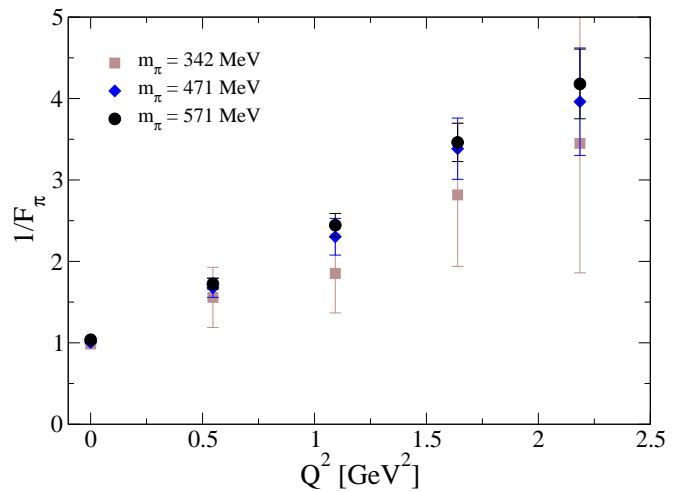} 
\end{center}
\caption{The inverse vector form factor for the three available masses 
on the $16^3\times 32$ lattice.}
\label{fig:InvFpi}
\end{figure}

Some combinations of initial and final momenta induce more statistical
noise than others. To increase our statistics we average over all
combinations that have the same momentum transfer. In particular, for
$Q^2=2\,p_0^2$ and $Q^2=6\,p_0^2$ we have four times as many data, and for
$Q^2=4\,p_0^2$ the averaging doubles our statistics.

When comparing the results to physical, renormalized quantities we have to
multiply with renormalization factors relating the raw results to the 
$\overline{\textrm{MS}}$-scheme. For the chirally improved Dirac operator 
these have been determined in \cite{GaGoHu04} and all results we show, except 
for the plots of the ratios, are already converted to the 
$\overline{\textrm{MS}}$-scheme. We always use values as 
determined in the chiral limit.

\subsection{Vector form factor}

Although the vector current is pointlike
and not conserved, the value $Z_V=0.9586(2)$ turns out to be close to
unity. This number converts the lattice bare results at $a =$ 0.148~fm to
the corresponding renormalized continuum quantities in the $\overline{\rm
MS}$ scheme at the scale of 2~GeV. The fact that the renormalization
constant $Z_V$ is so close to 1 means that we have better control over
this kind of systematic effects. 

The resulting $F_\pi (0)$ is not constrained to unit value, but comes
very close. This fact is obvious from the results summarized in Table
\ref{tab:fpi_vect}. These values have been computed by combining the
plateau mean values as discussed before; these have been obtained by
averaging the values at $t=3$ and 4 of the l.h.s. plateau and the values
at $t=20-23$ of the r.h.s plateau. The errors have been determined with
the jackknife method. The larger error bars in some of the plots are an
indication of the rather poor signal-to-noise ratio for some timeslices
close to the lattice center. These timeslices are however outside the
regions that we use for the fits. As already mentioned in 
Sect.\ \ref{sec:strategy}, we profit from the cancellation of kinematical 
factors by following the approach suggested in Ref.\ \cite{HeKoLa04}.

Fig. \ref{fig_fpi_q2} compares our results for the vector form factor 
with results from experiments \cite{Vo01,Be78,Am86,Br79}. In Fig.
\ref{fig_fpi_all} we show these results together with those for the small
lattice size $12^3\times 24$. For this smaller lattice size the
statistics was sufficient only for the data at $a\,m=0.06$. The
resulting form factor values are at other momentum transfer values than
those of the larger volume, but in reasonable agreement. 

In Fig.~\ref{fig:InvFpi} we plot the inverse of the vector form factor.
The electromagnetic pion form factor in the time-like region is dominated
by the $\rho$-meson. In the near space-like region it may therefore be
well approximated by a monopole form (\ref{eq:rvmd}) such that the
leading behavior of $1/F_\pi$ is linear. This is indeed observed in
Fig.~\ref{fig:InvFpi}. 

As the VMD model is just an approximation, one expects corrections due to
other resonances and more-particle channels in the time-like region. In
the fit these may be taken into account by further pole terms. Our data
shown in Fig.~\ref{fig:InvFpi} do not really require such a
multi-parameter fit and we therefore discuss only the results of a linear
fit.

\begin{figure}[tb]
\begin{center}
\includegraphics*[height=5cm]{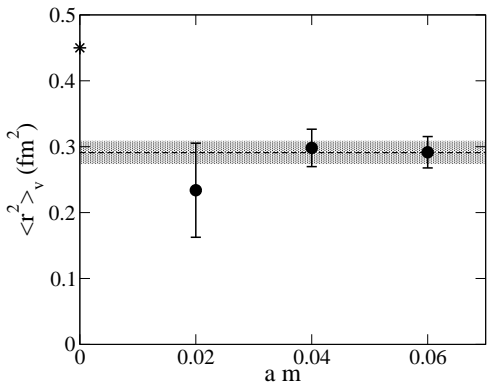}
\end{center}
\caption{The mean square radius for the vector form factor
as obtained for the three available masses 
on the $16^3\times 32$ lattice.
The asterisk denotes the average experimental value \cite{PDG04}, the error 
band shows the result of a constant fit to the three points.}
\label{fig:pionchargeradius}
\end{figure}

The derivative of the form factor at $Q^2=0$ gives the charge radius, as
shown in Fig.~\ref{fig:pionchargeradius}. The derivative has been
obtained from the linear fit to the inverse form factor. Whereas the
large mass results are compatible with a constant in $a\,m$, the number
for the lowest mass is smaller but has a very large statistical error. 
These values are quite compatible with numbers from \cite{HeKoLa04} at
comparable quark masses.

Obviously in our results there is an unknown systematic error due to the
quenched approximation. We refrain from applying QChPT extrapolations because 
the data are not accurate enough to reliably determine the unknown expansion 
parameters for the non-linear terms. Computing the average gives 
$\langle r^2\rangle_\textrm{\scriptsize v}=0.291(18)\;\textrm{fm}^2$. 

The value of $\langle r^2\rangle_\textrm{\scriptsize v}$ in the VMD model 
is inversely proportional to the mass squared of the $\rho$-meson. We obtain 
$m_\textrm{\scriptsize VMD}^2=0.80(5)\;\textrm{GeV}^2$. This agrees with the 
range of values obtained for $m_\rho^2$ in the analysis of 
Ref.~\cite{GaGoHa03a}
in the direct $\rho$-channel (e.g., $m_\rho^2=0.80\;\textrm{GeV}^2$ at
$a\,m=0.04$). These large values for the $\rho$-meson mass might at least in
part explain (via the VMD model) our low (as compared to experiment)
result for $\langle r^2\rangle_\textrm{\scriptsize v}$.

Like most other calculations we could afford only to work for a single value 
of the lattice spacing and therefore we have no control on scaling violations. 
However, due to the (approximate) GW nature of the CI operator we expect 
$\mathcal{O}(a)$ corrections to be very small \cite{GaGoHa03a}.

Due to the chosen method we have fewer momentum transfer values than
\cite{BoEdFl05}, who also work at larger lattices and at slightly smaller
quark masses. Our results, obtained with different action and Dirac operator, 
are in generally good agreement with the findings of other lattice calculations
in the discussed range of values for mass and momentum transfer
\cite{BoEdFl05,HeKoLa04}, even for those with dynamical background
\cite{BoEdFl05}. Higher quark masses overestimate the values of the form
factor, as it is expected due to the corresponding higher vector meson
mass. This then leads to an underestimation of the charge radius.

\subsection{Scalar form factor}

\begin{figure}[t]
\begin{center}
\includegraphics*[width=87mm]{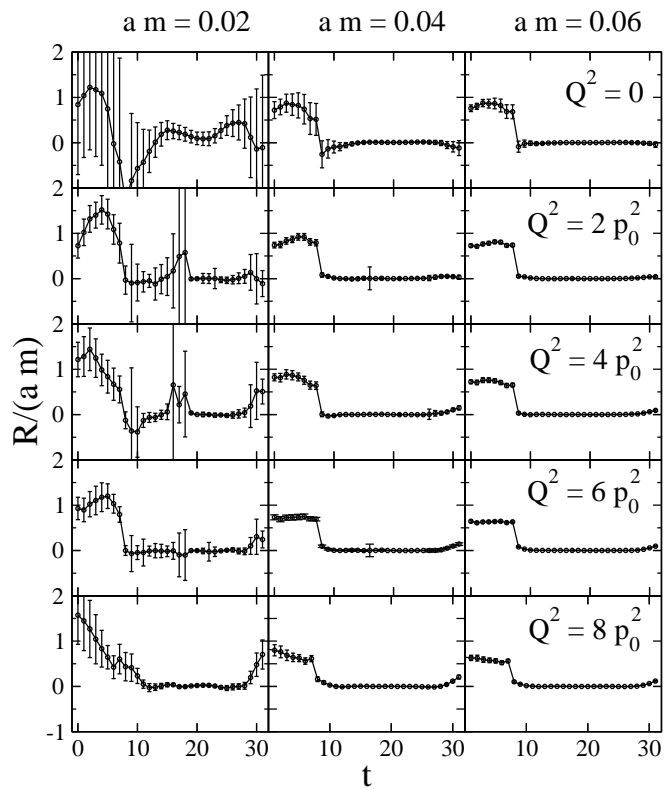} 
\end{center}
\caption{The ratio (\ref{eqcorr}) for the matrix element 
(\ref{scalarmatrixelement}) leading to the scalar form factor for all
quark masses and transfer momenta considered (lattice size $16^3\times 32$).}
\label{fig_rat_scal}
\end{figure}

\begin{table}[t]
\caption{Values of the pion scalar form factor (lattice size $16^3\times32$,
$a=0.148$~fm, without disconnected contributions)
obtained for different quark mass values.}
\label{tab:fpi_scal}
\begin{ruledtabular}
\begin{tabular}{cccc}
~\vspace{-8pt}\\
  $ Q^2$[GeV]  &     $a\, m$=0.02&  $a \,m$=0.04&   $a \,m$=0.06\\
\hline
   0   & 1.77(2.20)  & 1.30(39) &  1.42(17)\\
 0.546 & 2.09(53)    & 1.38(13) &  1.31(6) \\
 1.093 & 1.60(51)    & 1.31(13) &  1.24(7) \\
 1.639 & 1.61(56)    & 1.13(13) &  1.04(7) \\
 2.186 & 1.37(69)    & 1.03(16) &  0.96(9) \\
\end{tabular}
\end{ruledtabular}
\end{table}

The scalar form factor also has disconnected contributions which we
disregard here (like it is often done due to the inherent technical
complications of backtracking loops). We can notice from the upper plots
in Fig.\ \ref{fig_rat_scal} and the upper line in Table
\ref{tab:fpi_scal} that the scalar form factor at zero momentum transfer
divided by the quark mass has nearly the same value for all masses that
we have simulated. This means that $\Gamma_\pi (0)\sim M_\pi^2 \sim m$, 
as expected. We may determine the scalar radius squared from a
fit to the $Q^2$-dependence of the scalar form factor (now explicitly
normalized according to Eq.\ (\ref{defscalarradius})). 

\begin{figure}[t]
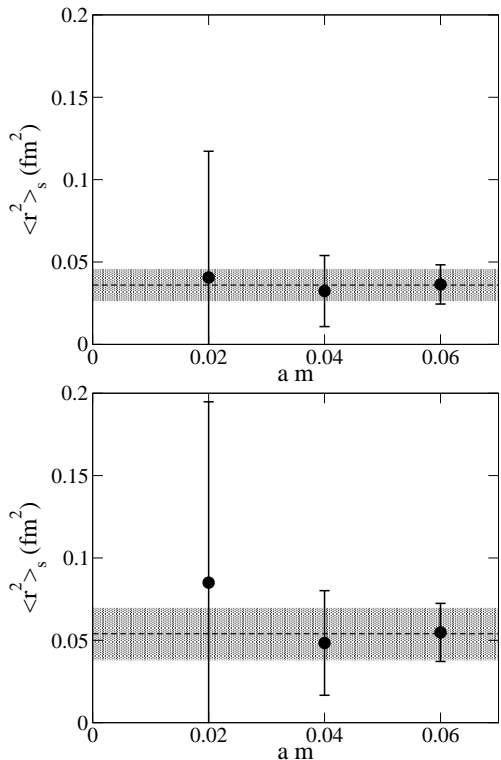

\begin{center}
\includegraphics*[height=5cm]{figs/PionScalarRadiusPlat.eps}\\
\includegraphics*[height=5cm]{figs/PionScalarRadiusInvPlat.eps}
\end{center}
\caption{The mean square radius for the three available masses 
on the $16^3\times 32$ lattice for the scalar form factor.
Upper plot: Determination from the linear fit to the values of
$\Gamma_\pi(Q^2)/\Gamma_\pi(0)$. Lower plot: Determination 
from the linear fit to the inverse of that ratio.}
\label{fig:pionscalarradius}
\end{figure}

In analogy to the vector form factor we expect that the scalar form factor
in the space-like region is a decreasing but positive definite function.
Thus a linear fit to its inverse seems appropriate. However, the result
for the scalar radius squared is quite sensitive to this assumption. In
Fig.\ \ref{fig:pionscalarradius} we compare the results of such a fit 
with those from a direct linear fit to our data in the space-like region 
and find a difference of almost 50\%. In either case, the resulting value 
may then be extrapolated to the chiral limit.

ChPT relates the pion decay constant to the scalar form factor radius via
\begin{equation}\label{eq:chptfpiscalrradius}
f_\pi/f=1 + \frac{1}{6}\langle r^2\rangle_s M_\pi^2 
+ \frac{13}{12}\xi+\mathcal{O}(\xi^2) \;,
\end{equation}
with
\begin{equation}
\xi =  \left( \frac{M_\pi}{4\,\pi\,f_\pi}\right)^2 \;.
\end{equation}
The chiral expansion of the pion decay constant should behave like
\cite{CoGaLe01}
\begin{equation}\label{eq:chptfpiexpansion}
f_\pi/f    =  1+\xi\,\bar \ell_4+\mathcal{O}(\xi^2) \;,
\end{equation}
where the value $\bar \ell_4=-\ln (M_\pi^2/ \Lambda^2)$ depends on the 
intrinsic QCD scale $\Lambda$ and in \cite{CoGaLe01} it is suggested to use
$\Lambda\approx 4\,\pi\,f_\pi$; in Ref.\;\cite{AnCaCo04} a value of 
$\bar \ell_4\approx 4.0\pm 0.6$ is quoted.
Eq.\ (\ref{eq:chptfpiscalrradius}) may be translated to
\begin{equation}
\langle r^2\rangle_s=\frac{3}{8\,\pi^2\,f_\pi^2}\,
\left(\bar \ell_4 - \frac{13}{12}+\mathcal{O}(\xi)\right)\;.
\end{equation}
Recent values quoted in that context are $\langle
r^2\rangle_s=0.61(4)\,\textrm{fm}^2$ \cite{AnCaCo04} or 
$0.75(7)\,\textrm{fm}^2$ \cite{Yn05}.

In QCD one expects correction terms with a logarithmic singularity in the 
valence quark mass $m$. As pointed out in \cite{Sh92}, the leading order
logarithmic term $m\, \log m$ of ChPT involves quark loops that are
however absent in the quenched case. There will be non-leading
logarithmic terms, though. One therefore could allow a term $m^2 \,\log
m$ in the extrapolating fit. In the range of mass values studied here,
the statistical accuracy of the data is too poor to render such an
extrapolation significant. We therefore exhibit only the results of a
constant extrapolation.

The resulting values (Fig.~\ref{fig:pionscalarradius}) are almost an order
of magnitude smaller than for the electromagnetic case. From the data in
Table \ref{tab:fpi_scal}, which takes into account the renormalization of
the scalar operator, $Z_S = 1.1309(9)$, as well as the kinematical
factors, we obtain an average for the scalar radius of $\langle
r^2\rangle_s=0.054(16)\;\textrm{fm}^2$ (based on the linear fit to the
inverse form factor). This is also smaller than the values expected for
full QCD. Results for the pion decay constant in a recent full QCD
lattice calculation \cite{AuBeDe04} via (\ref{eq:chptfpiscalrradius})
lead to the value $\langle r^2 \rangle_s = 0.5 \pm 0.1~\textrm{fm}^2$. A
corresponding analysis of the quenched BGR data \cite{GaHuLa05} gives
again a small value, $0.08-0.13\;\textrm{fm}^2$. This seems to imply that
$\langle r^2 \rangle_s$ is very sensitive to quenching and possibly to the
omission of the disconnected pieces. 

A recent unquenched computation with $N_f=2$ clover fermions \cite{HaAoFu05}, 
which uses matrix elements of the scalar current and not the pion 
decay constant, quotes a larger value for the scalar radius squared, 
$\langle r^2\rangle_s=0.60(15) \;\textrm{fm}^2$. However, this number results 
from a ChPT-motivated extrapolation to data that is actually of smaller size 
and for pion masses $\geq$ 550 MeV. 

\section{Conclusions}

Chirally improved fermions provide a framework for a first-principles 
lattice study of the hadron structure at comparatively small pion masses. 
In that context we present here our investigation of the vector and scalar
form factors of the pion. We combine the method of \cite{HeKoLa04} 
(which for the vector form factor avoids calculation of the pion energy and 
thus removes one source of statistical error) with a GW type action, that 
allows one to reach small pion masses for comparatively small lattice size.

The vector form factor describes the electromagnetic 
structure and our results are generally consistent with expectations of its 
general form. In view of the quenched approximation we expect deviations 
from the experimental numbers. The vector charge radius indeed comes out 40\%
smaller than in experiment and the scalar radius is much lower than one
would expect from unquenched calculations. 

Further progress could be achieved - as usual - by using better
statistics, more momenta, and larger lattices, both in lattice units (to
study finite volume effects) and in physical units (to access lower
transferred momenta).

\begin{acknowledgments}

Most computations were done on the Hitachi SR8000-F1 at the Leibniz 
Rechenzentrum in Munich, and we thank the staff for support.
S.~C. is supported by Fonds zur F\"orderung der Wissenschaftlichen Forschung 
in \"Osterreich, Project P16310-N08.
\end{acknowledgments}



\end{document}